\documentclass{PoS}

\usepackage{amsmath}
\usepackage{soul}
\title{Towards an extraction of $\hat{q}$ with an uncertainty
controlled energy loss Monte-Carlo}

\ShortTitle{Towards a precision $\hat{q}$ extraction}

\author{\speaker{Weiyao Ke}\\
        Duke University, Durham, North Carolina 27708 USA\\
        E-mail: \email{wk42@phy.duke.edu}}

\author{Yingru Xu\\
        Duke University, Durham, North Carolina 27708 USA\\
        E-mail: \email{yx59@phy.duke.edu}}
\author{Steffen A. Bass\\
        Duke University, Durham, North Carolina 27708 USA\\
        E-mail: \email{bass@phy.duke.edu}}

\abstract{Recent progress in open-heavy-flavor measurements and future experimental upgrades are bringing heavy-flavor physics into the precision era, allowing for strong quantitative constraints on
the transport properties of heavy quarks in the quark-gluon plasma. 
Starting from the LIDO transport model, which combines a matrix-element based linearized-Boltzmann transport and diffusion based Langevin equation, we have made two essential improvements to increase both the flexibility and physical accuracy of the model.
First, we have absorbed the pQCD scatterings with small momentum transfers to the medium into the diffusion part of the LIDO model, and we have restricted the use of vacuum matrix elements to large momentum transfer processes. This study allows us to construct a model that smoothly interpolates between a pure pQCD based approach and a radiation-improved Langevin equation by tuning a single scale parameter. 
Second, the Monte-Carlo implementation of the Landau-Pomeranchuk-Migdal effect of the original model has been improved to account for multiple scatterings of gluons. 
The simulated radiative energy loss can be tuned to quantitatively agree with semi-analytic theory calculations both for a static  (finite / infinite) medium and for a dynamic expanding medium.
With such improvements, the LIDO model will greatly facilitate the extraction of heavy-quark transport coefficient from a systematic model-to-data comparison.
}

\FullConference{International Conference on Hard and Electromagnetic Probes of High-Energy Nuclear Collisions\\
30 September - 5 October 2018\\
Aix-Les-Bains, Savoie, France}

\begin{document}

\section{Introduction}
Jet and heavy flavor probes provide unique information of the quark-gluon plasma (QGP). 
The jet transport coefficient $\hat{q}$ is often used to quantify the interaction between hard partons and the QGP medium, defined as the momentum broadening per unit time transverse to the direction of motion.
Generally, $\hat{q}$ can be both temperature- and momentum-dependent, and the determination of its functional form in terms of temperature and momentum is of great theoretical importance.
However, theoretical calculations of this dynamical quantity have been proven to be extremely hard using either perturbative techniques or Lattice simulations.
Instead, one seeks an alternative way to extract this basic quantity from experiments through a systematic model-to-data comparison.
Such a comparison requires both high quality measurements and a reliable modeling of the relevant physical processes with quantified theoretical uncertainty.
In this contribution, we describe a step forward in gauging the dynamical modeling of partonic transport inside QGP using existing theoretical calculations.
This calculation is performed using the framework of the LIDO transport model originally developed for heavy quarks \cite{Ke:2018tsh}.
Two essential physical improvements were made \cite{Ke:2018jem} that considerably increase the flexibility of the model and its physical accuracy, making it a compelling candidate to be used for a reliable extraction of the transport coefficient. 

\section{The Improved LIDO model}
The original LIDO model includes both elastic and inelastic scattering channels for the heavy quarks, solved by a linearized Boltzmann equation with perturbative matrix-elements.
In particular, the $2\rightarrow 3$ gluon radiation process is accompanied by a $3\rightarrow 2$ absorption process to maintain detailed balance. 
Between perturbative scatterings, heavy quarks undergo diffusive motion with an empirical transport coefficient $\hat{q}_{\textrm{NP}}$ that mimics any non-perturbative effects.

Recently, we made two major improvements \cite{Ke:2018jem} that are described as follows. 
First, the perturbative interactions between the hard partons and the medium are further divided into hard- and soft-modes by a separation scale $Q_{\textrm{cut}}$ that is proportional to the screening mass $m_D$.
This separation was originally proposed in \cite{Ghiglieri:2015ala} to facilitate next-to-leading-order jet transport calculations, while its advantage for this work is to allow for a separate and proper modeling of each regime.
The soft interactions whose associated momentum transfers are smaller than $Q_{\textrm{cut}}$ are approximated by a diffusion process, absorbed into the LIDO diffusion sector.
This diffusive motion also induces an effective $1\rightarrow 2$ gluon radiation process.
Interactions with momentum transfers larger than $Q_{\textrm{cut}}$ are treated as scatterings using vacuum matrix-elements.
Now, the transport coefficient $\hat{q}$ receives contributions from soft, hard, and non-perturbative interactions,
\begin{eqnarray}
\hat{q} = \hat{q}_\textrm{S} + \hat{q}_{\textrm{H}} + \hat{q}_{\textrm{NP}},
\end{eqnarray}
Where the hard contribution can be obtained by numerically integrating the elastic collision rate,
\begin{equation}
\hat{q}_{\textrm{H}} = \frac{1}{2E}\int\frac{dp'^3}{(2\pi)^3 2 E'}f_0(p')2(s-M^2)\int_{-(s-M^2)^2/s}^{-Q_{\textrm{cut}}^2} q^2 \frac{d\sigma_{\textrm{el}}}{dt} dt.
\end{equation}
The soft contribution is obtained as \cite{Ghiglieri:2015ala},
\begin{equation}
\hat{q}_{\textrm{S}} = C_A T \int_0^{Q_{\textrm{cut}^2}} \alpha_s(q^2) \frac{m_D^2 dq^2}{q^2+m_D^2}
\end{equation}
Both $\hat{q}_\textrm{S}$ and $\hat{q}_{\textrm{H}}$ depend on a medium scale parameter $\mu$ that enters in the evaluation of the coupling constant by introducing a bound for the running scale at $\mu\pi T$, $\alpha_s(Q) = \alpha_s(\max\{Q, \mu\pi T\})$. They also depend on the separation scale $Q_{\textrm{cut}}$ individually, but their sum is approximately $Q_{\textrm{cut}}$ independent.
Finally, $\hat{q}_{\textrm{NP}}$ is a parametric function accounting for non-perturbative effects, which we will set to zero for the rest of this proceeding.

\begin{figure}
\centering
\includegraphics[width=.63\textwidth]{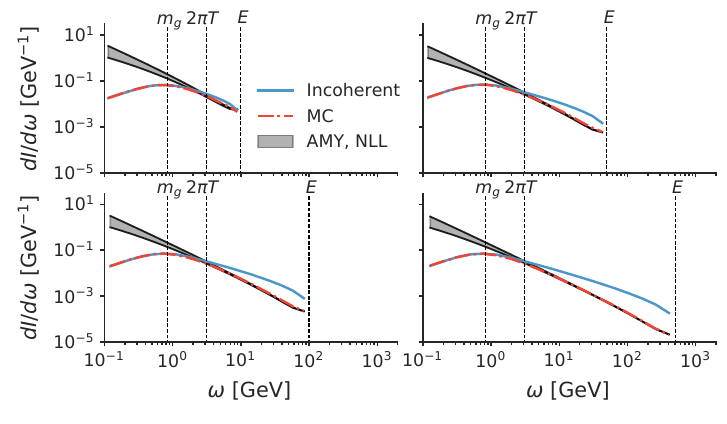}
\includegraphics[width=.36\textwidth]{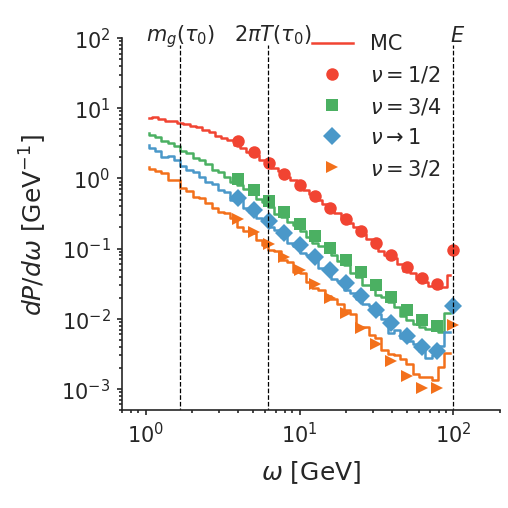}
\caption{Left: comparing simulated gluon radiation spectra from a light quark passing through an infinite medium to the AMY-NLL calculation \cite{Arnold:2008zu}. Right: comparing simulated gluon radiation in an expanding medium to the BDMPS calculations (symbols) \cite{Baier:1998yf}. }\label{fig:compare}
\end{figure}

Another improvement is a more sophisticated Monte-Carlo implementation of the QCD analog of the Landau-Pomeranchuk-Migdal (LPM) effect \cite{Baier:1996kr,Zakharov:1996fv}. 
The LPM effect strongly suppresses the radiative energy loss compared to the incoherent calculation.
Therefore, a meaningful parameter extraction from a model-to-data comparison requires a reliable implementation of this effect.
Our strategy is motivated by the approach described in \cite{Zapp:2011ya}.
We first generate gluon radiations within each time-step according to the incoherent $2\rightarrow 3$ and $1\rightarrow 2$ radiation rates.
These gluons, termed as ``preformed-gluons", are not regarded as independent objects, but are associated with their mother parton.
A mother parton may carry an arbitrary number of preformed-gluons, and they only interact via diffusion or elastic scatterings during their initial evolution.
This evolution continues until the time $t$ when its formation time $\tau_f\sim \omega/k_\perp^2$ satisfies $\tau_f < t-t_0$, where $t_0$ is the time when the preformed-gluon is created.
Then, this gluon is considered as ``formed" with a probability $P = \tilde{\lambda}/\tau_f$ and carries away energy and momentum from the mother parton; otherwise, it is discarded without causing energy loss of the mother parton.
Here $\tilde{\lambda}\sim m_D^2/\hat{q}$ is an effective mean-free-path that is well defined for both diffusion and scattering processes.

This procedure is validated by tuning the relation $\tilde{\lambda}\propto m_D^2/\hat{q}$ so that the simulated radiation spectra agree with  semi-analytic calculations. 
In Fig. \ref{fig:compare} (left), simulations using $\alpha_s = 0.3$ in an infinite medium with temperature $T=0.5$ GeV quantitatively agree with the next-to-leading-log calculations of the AMY formula \cite{Arnold:2008zu} in the LPM regime ($2\pi T \lesssim \omega < E$).
In the Bethe-Heitler region $\omega \lesssim 2\pi T$, the results converge to those obtained from incoherent simulations.
Moreover, quite remarkably, this approach also applies to an expanding medium which is a vital for using a realistic medium background obtained from hydrodynamic calculation.
In Fig. \ref{fig:compare} (right), we compared the simulations to the BDMPS calculation in an expanding medium with a power-law temperature profile $T^3 \propto \tau^{1/\nu-2}$ \cite{Baier:1998yf}. 
Only the diffusion sector and associated diffusion-induced radiation are turned on in this comparison, so that both the LIDO model and theoretical formula use a same input of $\hat{q}_g = C_A \alpha_s T m_D^2$. 
The model nicely describes the radiation spectra at different medium expansion rates $\nu =$ 0.5 (static), 0.75, 1.0 (Bjorken), and 1.5. 
These agreements with theoretical expectations demonstrate that the model has reached a precision level in reflecting the underlying physics (leading-order calculation).
Please refer to \cite{Ke:2018jem} for more comprehensive comparisons to theory.

\begin{figure}
\centering
\includegraphics[width=.8\textwidth]{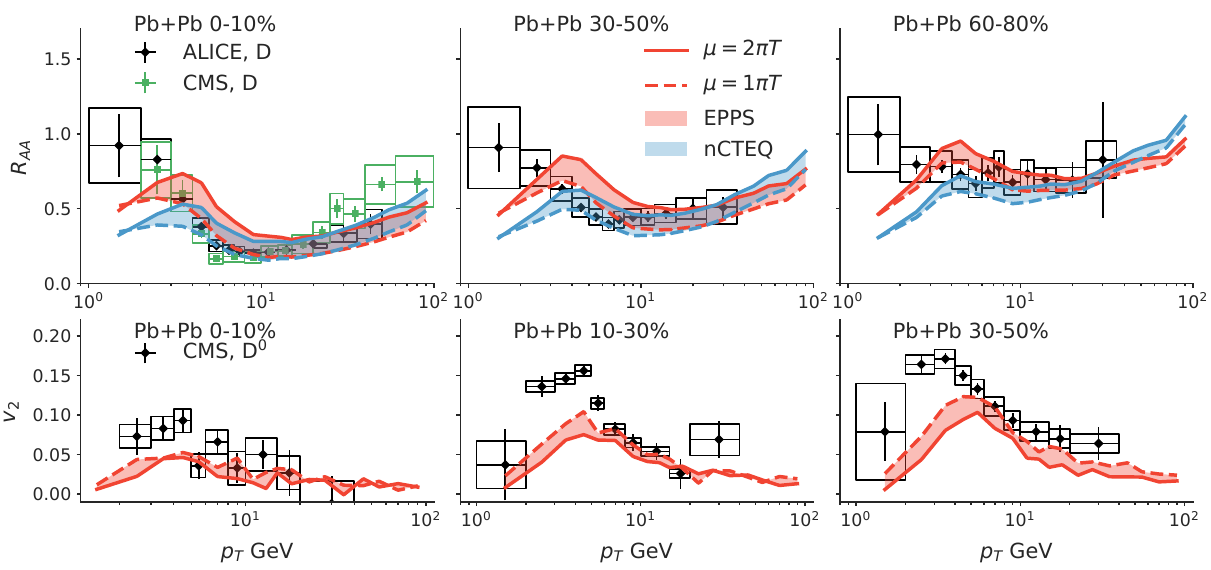}
\caption{Calculation of $D$-meson nuclear modification factor and momentum anisotropy in Pb+Pb collisions at 5.02 TeV. This calculation includes only perturbative contribution and the bands denote uncertainty in the choice of the medium scale parameter.}\label{fig:benchmark}
\end{figure}

\section{Benchmark Predictions and Future Perspectives}
Before applying the model to the extraction of the heavy quark transport coefficient $\hat{q}$ using a sophisticated statistical analysis, it is instructive to ``predict" observables with a few reasonable choices of the model parameters.
For this purpose, we retain the perturbative contribution ($\hat{q}_{\textrm{NP}} = 0$). 
Therefore, the only physical parameter of the model is the medium scale $\mu\pi T$ that appears in the coupling constant $\alpha_s(\max\{Q, \mu\pi T\})$.
Using the leading order running coupling, $\hat{q}_{\textrm{S}}$ used in the diffusion sector can be integrated approximately as,
\begin{eqnarray}
\frac{\hat{q}_S}{6\pi C_R\alpha_s^2(\mu\pi T)T^3} \approx
\begin{cases}
 \ln\left(1+Q_{\textrm{cut}}^2/m_D^2\right), \textrm{ for } Q_{\textrm{cut}} < \mu\pi T, \\
 \\
\ln\left[\left(1+\frac{(\mu\pi T)^2}{m_D^2}\right)\frac{(\mu\pi T)^2}{\Lambda^2}\right] - \frac{2\ln^2(\mu\pi T/\Lambda)}{\ln[Q_{\textrm{cut}}/\Lambda]}, \textrm{ for } Q_{\textrm{cut}} > \mu\pi T.
\end{cases}
\end{eqnarray}
Here we show calculations for two natural choices of $\mu = 1$ and $\mu=2$.
The event-by-event hydrodynamic simulation is performed using the package and ``best-fit" parameters obtained in \cite{Bernhard:2018hnz}.
We then calculate the nuclear modification factor $R_{AA}$ and the momentum anisotropy $v_2$ of $D$-meson at 5.02 TeV in Pb-Pb collisions at the LHC.
The results are compared to measurements by the ALICE collaboration \cite{Acharya:2018hre} and the CMS collaboration \cite{Sirunyan:2017plt}.
We find that these natural choices of $\mu$ reasonably describe the data at $p_T > 10$ GeV.
At low $p_T$, a model that implements only perturbative processes will fail to describe the observed large momentum anisotropy. 
This discrepancy may be resolved by the calibration of model parameters that include both perturbative and parametrized non-perturbative contributions to the transport coefficient.

\section{Summary}
In summary, we have presented an improved version of the LIDO model that 1) divides probe-medium interactions into soft- and hard-modes to allow for their separate treatment and 2) contains an improved treatment of the LPM effect tuned to quantitatively agree with theory calculations.
Without a full calibration of model parameters to available data, a natural choice of model parameter already shows significant promise in describing high-$p_T$ data.
Using this model, a large scale Bayesian calibration for the extraction of the functional form of the heavy quark transport coefficient will be explored in the future.

\begin{acknowledgments}
SAB, YX, and WK  are supported by the U.S. Department of Energy Grant no. DE-FG02-05ER41367.
SAB and WK are also funded by NSF grant ACI-1550225. The authors thank Jean-Fran\c{c}ois Paquet, Tianyu Dai and Wenkai Fan for helpful discussions.
\end{acknowledgments}

\bibliographystyle{JHEP}
\bibliography{ref}

\end{document}